# Deep Sparse Coding for Non-Intrusive Load Monitoring

Shikha Singh and Angshul Majumdar, *Senior Member IEEE*

*Abstract—* Energy disaggregation is the task of segregating the aggregate energy of the entire building (as logged by the smartmeter) into the energy consumed by individual appliances. This is a single channel (the only channel being the smart-meter) blind source (different electrical appliances) separation problem. The traditional way to address this is via stochastic finite state machines (e.g. Factorial Hidden Markov Model). In recent times dictionary learning based approaches have shown promise in addressing the disaggregation problem. The usual technique is to learn a dictionary for every device and use the learnt dictionaries as basis for blind source separation during disaggregation. Prior studies in this area are shallow learning techniques, i.e. they learn a single layer of dictionary for every device. In this work, we propose a deep learning approach – instead of learning one level of dictionary, we learn multiple levels of dictionaries for each device. These multi-level dictionaries are used as a basis for source separation during disaggregation. Results on two benchmark datasets show that our method outperforms state-of-the-art techniques.

## I. Introduction

ENERGY disaggregation, is the task of segregating the combined energy signal of a building into the energy consumption of individual appliances. Currently, residential and commercial buildings account for 40% of total energy consumption [1], and studies have estimated that 20% of this consumption could be avoided with improvement in user behavior [2]. Disaggregation presents a way in which consumption patterns of individuals can be learned by the utility company. This information would allow the utility to present this information to the consumer, with the goal of increasing consumer awareness about energy usage. Studies have shown that this is sufficient to improve consumption patterns [3].

The approach towards energy disaggregation is broadly based on the nature of the targeted household and commercial appliances. These appliances can be broadly categorised as simple two-state (on/off) appliances such as electrical toasters and irons; more complex multistate appliances like refrigerators and washing machines; and continuously varying appliances such as IT loads (printers, modems, laptops etc.). The earliest techniques were based on using real and reactive power measured by residential smart meters. The appliances' power consumption patterns were modelled as finite state machines [4]. These techniques were successful for disaggregating simple two state and multistate appliances, but they performed poorly in the case of time-varying appliances which do not show a marked step increase in the power. Even in recent times, there are techniques that primarily disaggregate based on jumps and drops in the power signature [5, 6].

More recent techniques, based on stochastic finite state machines (Hidden Markov Models) [7], have improved upon the prior approach. Another approach is based on learning a basis for individual appliances. Sparse coding and dictionary learning based approaches like [8] fall under this category. A recent study introduced the powerlet technique to learn energy signatures [9]. Given the limitations in space it is not possible to discuss all the prior studies in this area in detail; the interested reader should peruse [10].

The success of deep learning over the past decade is a common knowledge. In this work, we give an alternate interpretation to sparse coding / dictionary learning – we show a relationship between dictionary learning and neural networks. Then we will show how the sparse coding approach can be extended to deeper architectures; thereby leading to deep sparse coding.

In the sparse coding approach introduced in this context by [8], the idea is to learn a basis for each electrical appliance from training data. During disaggregation, the combined power (from several appliances) is assumed to be a superposition of the powers from individual appliances, and is expressed in terms of the learned basis. By estimating the loading coefficients, it is possible to calculate how much power was consumed by each appliance. Our basic extension is simple. Instead of learning a single level of basis / dictionary we learn multiple layers – motivated by deep learning in other areas. The concatenated multi-layered basis is used for signal disaggregation.

It must be noted that our work is not related to hierarchical / structured dictionary learning techniques [11-13]; although the title of [13] carries the terms 'deep', 'sparse' and 'coding' – it is basically a hierarchical approach; not a deep one. Hierarchical learning is a shallow (single level) learning technique, where a single level of dictionary is learnt, but the dictionary atoms maintain a hierarchical structure. It is similar to 'learning' a wavelet like decomposition for 'tree-structured' sparsity on any piecewise smooth signal.

Experimental results are carried out on two benchmark datasets – REDD and Pecan Street. We show that our proposed simple extension achieves better performance than state-of-the-art shallow architectures.

## II. Literature Review

### A. Deep Learning

Deep learning (stacked autoencoder and deep belief network) and dictionary learning fall under the purview of representation learning. However, the relationship between them are not well explored.

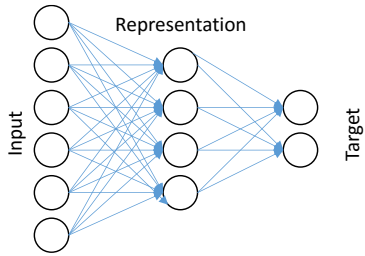

Fig. 1. Single Representation Layer Neural Network

Fig. 1 shows the diagram of a simple neural network with one representation (hidden) layer. The problem is to learn the network weights between the input and the representation and between the representation and the target. This can be thought of as a segregated problem (see Fig. 2).

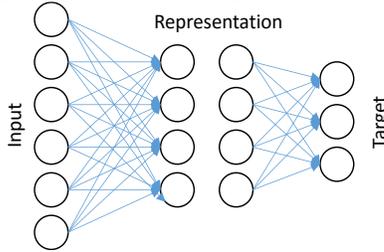

Fig. 2. Segregating the Neural Network

Learning the mapping between the representation and the target is straightforward. The challenge is to learn the network weights (from input) and the representation. Broadly speaking this is the topic of representation learning.

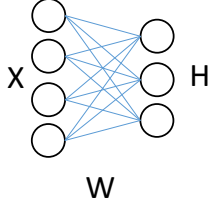

Fig. 3. Restricted Boltzmann Machine

Restricted Boltzmann Machine (RBM) [14] is one technique to learn the representation layer. The objective is to learn the network weights (W) and the representation (H). This is achieved by optimizing the Boltzman cost function given by:

$$p(W,H) = e^{H^T W X} \qquad (1)$$

Basically RBM learns the network weights and the representation / feature by maximizing the similarity between the projection of the input and the features in a probabilistic sense. Since the usual constraints of probability apply, degenerate solutions are prevented. The traditional RBM is restrictive – it can handle only binary data. The Gaussian-Bernoulli RBM [15] partically overcomes this limitation and can handle real values between 0 and 1. However, it cannot handle arbitrary valued inputs (real or complex).

Deep Boltzmann Machines (DBM) [16] is an extension of RBM by stacking multiple hidden layers on top of each other (Fig. 2). The RBM and DBM are undirected graphical models. For training deep architectures, targets are attached to the final layer and fine-tuned with back propagation.

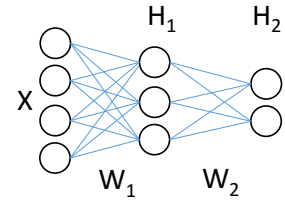

Fig. 4. Deep Botlzmann Machine

The other prevalent technique to train the representation layer of a neural network is by autoencoder [17]. The architecture is shown in Fig. 4.

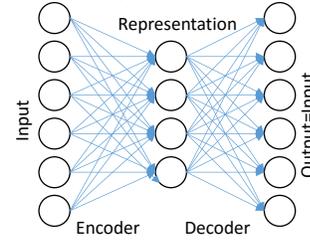

Fig. 5. Autoencoder

$$\min_{W,W'} \| X - W' \phi(WX) \|_F^2 \qquad (2)$$

The cost function for the autoencoder is expressed above. $W$ is the encoder, and $W'$ is the decoder. The activation function $\varphi$ is usually of tanh or sigmoid such that it squashes the input to normalized values. This prevents degeneracy in the solution. The autoencoder learns the encoder and decoder weights such that the reconstruction error is minimized. Essentially it learns the weights so that the representation $\phi(WX)$ retains almost all the information (in the Euclidean sense) of the data, so that it can be reconstructed back. Once the autoencoder is learnt, the decoder portion of the autoencoder is removed and the target is attached after the representation layer.

To learn multiple layers of representation, the autoencoders are nested into one another. This architecture is called stacked autoencoder.

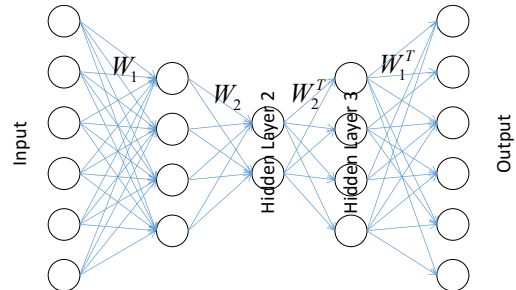

Fig. 6. Two-layer Stacked Autoencoder

For such a stacked autoencoder, the optimization problem is complicated.

$$\min_{W_1,W_2,W_1',W_2'} \| X - W_1' \varphi(W_2' \varphi(W_2 \varphi(W_1 X))) \|_F^2 \qquad (3)$$

The workaround is to learn the layers are learnt in a greedy fashion [18]. First the outer layers are learnt (see Fig. 7); and using the features from the outer layer as input for the inner layer, the weights for the inner layer are learnt.

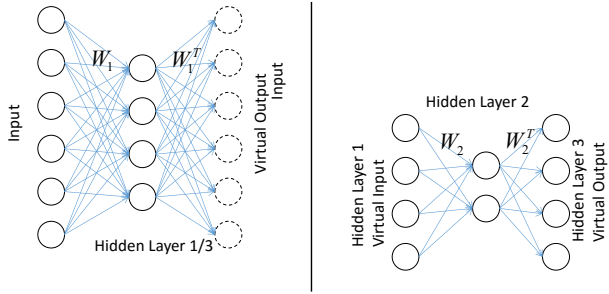

Fig. 7. Greedy Learning

For training deep neural networks, the decoder portion is removed and targets attached to the inner layer. The complete structure is then fine-tuned with backpropagation.

*B. Sparse Coding*

Kolter et al in [8], assumed that there is training data collected over time, where the smartmeter logs only consumption from a single device only. This can be expressed as $X_i$ where $i$ is the index for an appliance, the columns of $X_i$ are the readings over a period of time.
For each appliance they learnt a basis, i.e. they expressed:
$$X_i = D_i Z_i, \; i = 1...N \quad (3)$$
where $D_i$ represents the basis/dictionary, $Z_i$ are the loading coefficients, assumed to be sparse and $N$ is the total number of appliances.

This is a typical dictionary learning problem with sparse coefficients – there are several ways to solve (1). The most popular being the KSVD algorithm by [19]. However in [8] a more direct optimization based approach was formulated.
$$\min_{D_i, Z_i} \|X_i - D_i Z_i\|_F^2 + \lambda \|Z_i\|_1, \; i = 1...N \quad (4)$$
On top of (4), there is an additional constraint on the positivity of the loading coefficients to conform to physics.

The problem (40 is non-convex. It is solved via alternating minimization. In one step, the sparse coefficients ($Z$'s) are updated assuming the codebook / dictionary ($D$) to be fixed (5a); in the next stage, the codebook is updated assuming the coefficients to be constant (5b). During the sparse coding stage, the negative values in the sparse code are put to zero.
$$\min_{Z_i} \|X_i - D_i Z_i\|_F^2 + \lambda \|Z_i\|_1 \quad (5a)$$
$$\min_{D_i} \|X_i - D_i Z_i\|_F^2 \quad (5b)$$
In order to prevent degenerate solutions (where D is very large and Z is very small or vice versa) the dictionary atoms are normalized after every iteration.

During actual operation, several appliances are likely to be in use simultaneously. In such a case (assuming reactive loads only) the aggregate power read by the smartmeter is a sum of the powers for individual appliances. Thus if $X$ is the total power from N appliances (where the columns indicate smartmeter readings over the same period of time as in training) the aggregate power can be modeled as:
$$X = \sum_i X_i = \sum_i D_i Z_i \quad (6)$$

Given this model, it is possible to find out the loading coefficients of each device by solving the following sparse recovery problem,
$$\min_{Z_1,...,Z_N} \|X - [D_1 | ... | D_N]\|_F^2 + \lambda \left\| \begin{array}{c} Z_1 \\ ... \\ Z_N \end{array} \right\|_1 \quad (7)$$

Here a positivity constraint on the loading coefficients is enforced as well. This is a convex problem since the basis are fixed. Once the loading coefficients are estimates, one can easily compute the power consumption from individual devices
$$\hat{X}_i = D_i Z_i, \; i = 1...N \quad (8)$$
We have discussed the fundamental concept behind sparse coding based energy disaggregation. In [8] and [9] more sophisticated codebook learning techniques have been proposed with additional penalty terms. Owing to limitations in space, we cannot discuss them here; the interested reader may peruse the aforesaid papers. In this work, we will show that even without complicated penalties, we improve upon the state-of-the-art simply by learning deeper levels of dictionaries.

### III. PROPOSED DEEP SPARSE CODING

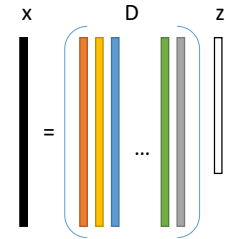

Fig. 8. Dictionary Learning

The popular interpretation for dictionary learning is that it learns a basis (D) for representing (Z) the data (X) (see Fig. 8); for sparse coding, the representation need be sparse. The columns of D are called 'atoms'. In this work, we have an alternate interpretation of dictionary learning. Instead of interpreting the columns as atoms, we can think of them as connections between the input and the representation layer (Fig. 9). To showcase the similarity, we have kept the color scheme intact in Fig. 8.

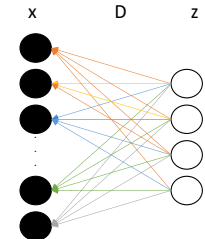

Fig. 9. Neural Network type Interpretation

Unlike a neural network which is directed from the input to the representation, the dictionary learning kind of network points in the other direction – from representation to the input. This is what is called 'synthesis dictionary learning' in signal processing. The dictionary is learnt so that the features (along with the dictionary) can synthesize / generate the data.

Till date dictionary learning / sparse coding had been a shallow architecture. The dictionary ($D_1$) is learnt such that the features (Z) synthesize the data along (X) with the dictionary. This is expressed as,

$$X = D_1 Z \qquad (9)$$

We propose to extend the shallow learning into multiple layers – leading to deep sparse coding. Mathematically, the representation at the second layer can be written as:

$$X = D_1 D_2 Z \qquad (10)$$

Note that it is not possible to collapse the two dictionaries $D_1 D_2$ (10) into a single level of dictionary ($D_1$) (9); the two formulations would not be equivalent. This is because (9) is a bi-linear problem whereas (10) is a tri-linear problem; therefore the features obtained from (9) would not be the same as those of (10) even if the dimensions match.

In (10) we show two levels of dictionaries; we can go deeper, to 3 and 4 layers; in that case deep dictionary learning can be expressed as (for N layers),

$$X = D_1 D_2 ... D_N Z \qquad (11)$$

There is no theoretical reason for finding deeper representations. However, proponents of deep learning argue that by finding deeper representations one can find more compact and abstract features that helps in the learning task. Usually there is a trade-off between going deeper and over-fitting. As one goes deeper, more and more parameters need to be learnt; thus the requirement for training data increases (leads to over-fitting). To prevent this one needs to find a compromise between abstraction and over-fitting.

There are two ways to solve it (11). The first one is a greedy approach. This is easy since the basic building blocks (shallow dictionary learning) are already available. But the limitation of this technique is that there is no feedback between the layers. The second solution (the exact solution) has not been hitherto solved. In this work we solve it variable splitting followed by alternating minimization. We will discuss both the solutions in the next two sub-sections.

*A. Greedy Solution*

This is the easier of the two solutions. Here, for the first layer, we express: $Z_1 = D_2 ... D_N Z$; so that the problem (11) can be formulated as,

$$X = D_1 Z_1 \qquad (12)$$

The coefficient $Z_1$ in the first layer is not sparse, hence the learning problem can be phrased as,

$$\min_{D_1, Z_1} \|X - D_1 Z_1\|_F^2 \qquad (13)$$

This is solved by alternating minimization.

$$Z_1 \leftarrow \min_{Z_1} \|X - D_1 Z_1\|_F^2 \qquad (14a)$$

$$D_1 \leftarrow \min_{D_1} \|X - D_1 Z_1\|_F^2 \qquad (14b)$$

Iterations are continued till local convergence.

In the second layer, we substitute $Z_2 = D_3 ... D_N Z$, leading to

$$Z_1 = D_2 Z_2 \qquad (15)$$

As before, this can be solved via alternating minimization. This can be continued till the last layer. At this layer, the formulation turns out to be,

$$Z_{N-1} = D_N Z \qquad (16)$$

Here, the coefficient needs to be sparse. Hence the alternating minimization turns out to be the same as sparse coding (5).

This is an easy approach. The basic building blocks for solving this problem are well studied. There are theoretical studies on single layer dictionary learning that prove optimality of alternating minimization regarding convergence (to local minima) [20-23]. But the problem with the greedy approach is that, information flows only in one direction, there is no feedback from latter layers to previous ones. Usually in deep learning, this issue is addressed by fine-tuning. However there is no scope of fine-tuning here since it is an unsupervised problem – there are no targets / outputs from which one can back-propagate.

*B. Exact Solution*

The goal is to solve (11). The exact solution is expressed as,

$$\min_{D_1, D_2, D_3, Z} \|X - D_1 D_2 ... D_N Z\|_F^2 + \lambda \|Z\|_1 \qquad (17)$$

An elegant way to address this problem is to use the Split Bregman approach [24]; variable splitting is a standard technique in signal processing these days [25-27]. We substitute $Y_1 = D_2 ... D_N Z$ and in order to enforce equality at convergence, introduce the Bregman relaxation variable ($B_1$). This leads to,

$$\min_{D_1, D_2, D_3, Z, Y_1} \|X - D_1 Y_1\|_F^2 \\ + \mu_1 \|Y_1 - D_2 ... D_N Z - B_1\|_F^2 + \lambda \|Z\|_1 \qquad (18)$$

To simplify (18) we substitute, $Y_2 = D_3 ... D_N Z$ and introduce another Bregman relaxation variable. This leads to,

$$\min_{D_1, D_2, D_3, Z, Y_1, Y_2} \|X - D_1 Y_1\|_F^2 + \mu_1 \|Y_1 - D_2 Y_2 - B_1\|_F^2 \\ + \mu_2 \|Y_2 - D_3 ... D_N Z - B_2\|_F^2 + \lambda \|Z\|_1 \qquad (19)$$

The process of substitution and introduction of Bregman variables can be continued till the last level. This leads to the following formulation,

$$\min_{D_1, D_2, D_3, Z, Y_1, Y_2, ..., Y_N} \|X - D_1 Y_1\|_F^2 + \mu_1 \|Y_1 - D_2 Y_2 - B_1\|_F^2 \\ + ... + \mu_{N-1} \|Y_{N-1} - D_N Z - B_{N-1}\|_F^2 + \lambda \|Z\|_1 \qquad (20)$$

Although this is not exactly a separable problem, we can use the method of alternating directions to break it down to several simpler sub-problems. Showing it for N levels is cumbersome, so we do it for 3 levels without loss of generality.

P1: $\min_{D_1} \|X - D_1 Y_1\|_F^2$

P2: $\min_{Y_1} \|X - D_1 Y_1\|_F^2 + \mu_1 \|Y_1 - D_2 Y_2 - B_1\|_F^2$

P3: $\min_{D_2} \|Y_1 - B_1 - D_2 Y_2\|_F^2$

P4: $\min_{Y_2} \mu_1 \|Y_1 - B_1 - D_2 Y_2\|_F^2 + \mu_2 \|Y_2 - D_3 Z - B_2\|_F^2$

P5: $\min_{D_3} \|Y_2 - B_2 - D_3 Z\|_F^2$

$$P6: \min_Z \mu_3 \|Y_2 - B_2 - D_3 Z\|_F^2 + \lambda \|Z\|_1$$

All the sub-problems, P1-P5, are linear least squares problems having a closed form solution. Therefore solving the sub-problems is straightforward. The last problem P6 is an $l_1$-minimization problem that can be solved efficiently using iterative soft thresholding [28].

In every iteration, the Bregman relaxation variable needs to be updated as follows,

$$B_1 \leftarrow Y_1 - D_2 Y_2 - B_1$$
$$B_2 \leftarrow Y_2 - D_3 Z - B_2$$

There are two stopping criteria for the Split Bregman algorithm. Iterations continue till the objective function converges (to a local minima). The other stopping criterion is a limit on the maximum number of iterations. We have kept it to be 200.

### C. Energy Disaggregation

In energy disaggregation by sparse coding, a codebook is learnt for every appliance [8] (3). The codebook learnt in prior studies are shallow. In this work, we propose to learn deep sparse codebook for every appliance; instead of (3) we will have for every appliance,

$$X^{(i)} = D_1^{(i)} D_2^{(i)} ... D_N^{(i)} Z \quad (21)$$

We have changed the notation a bit for ease of expression. The superscript denotes the $i^{th}$ appliance.

The codebook / dictionary for every appliance is learnt using the proposed technique (greedy or exact). Here we enforce the usual constraints – i) non-negativity of sparse coefficients, and ii) normalization of codebook.

Once the codebook for every appliance is learnt the disaggregation proceeds as before (7). The only difference between the previous shallow techniques and the proposed technique is that the codebook for each appliance is a cascade of codebooks / dictionaries – not a single one as in (6).

$$\min_{Z_1,...,Z_N} \left\| X - \left[ D^{(i)} | ... | D^{(N)} \right] \right\|_F^2 + \lambda \left\| \begin{matrix} Z_1 \\ ... \\ Z_N \end{matrix} \right\|_1 \quad (22)$$

where $D^{(i)} = D_1^{(i)} D_2^{(i)} ... D_N^{(i)}$.

Once the loading coefficients are solved for, the energy consumed by individual appliances is calculated as before, i.e. multiplying the cascaded codebook with the corresponding coefficients.

## IV. EXPERIMENTAL EVALUATION

In recent times, several research papers have been published proposing alternate signatures for load monitoring. In [29] a new current sensor is proposed. In [30], a derivative power signature is investigated for non-intrusive load monitoring. In a similar vein, [31] empirically tests a V-I trajectory based load signature. Even though the research on alternate signatures is promising, most studies on NILM depend on the standard smart-meter data for monitoring. Therefore in this work we will follow the same. We evaluate on two popular datasets – REDD and Pecan Street.

### A. REDD Dataset

We report results on two datasets. The first one is the REDD dataset [32] – a moderate size publicly available dataset for electricity disaggregation. The dataset consists of power consumption signals from six different houses, where for each house, the whole electricity consumption as well as electricity consumptions of about twenty different devices are recorded. The signals from each house are collected over a period of two weeks with a high frequency sampling rate of 15kHz. In the standard evaluation protocol, the 5[th] house is omitted since the data from this one is insufficient.

TABLE I
DESCRIPTION OF APPLIANCES IN HOUSES

| House | Appliances |
|---|---|
| 1 | Electronics, Lighting, Refrigerator, Disposal, Dishwasher, Furnace, Washer Dryer, Smoke Alarms, Bathroom GFI, Kitchen Outlets, Microwave |
| 2 | Lighting, Refrigerator, Dishwasher, Washer Dryer, Bathroom GFI, Kitchen Outlets, Oven, Microwave, Electric Heat, Stove |
| 3 | Electronics, Lighting, Refrigerator, Disposal, Dishwasher, Furnace, Washer Dryer, Bathroom GFI, Kitchen Outlets, Microwave, Electric Heat, Outdoor Outlets |
| 4 | Lighting, Dishwasher, Furnace, Washer Dryer, Smoke Alarms, Bathroom GFI, Kitchen Outlets, Stove, Disposal, Air Conditioning |
| 6 | Lighting, Refrigerator, Disposal, Dishwasher, Washer Dryer, Kitchen Outlets, Microwave, Stove |

The disaggregation accuracy is defined by [32] as follows,

$$Acc = 1 - \frac{\sum_t \sum_n \left| \hat{y}_t^{(i)} - y_t^{(i)} \right|}{2 \sum_t \bar{y}_t}$$

where $t$ denotes time instant and $n$ denotes a device; the 2 factor in the denominator is to discount the fact that the absolute value will "double count" errors. There may be other metrics for evaluating disaggregation results like precision, recall and F-measure or more recent measures proposed in [33], but disaggregation accuracy is still widely accepted and we continue using it here.

We compare the performance of our proposed method with the Factorial HMM (FHMM) based technique [32], Powerlet based Energy Disaggregation (PED) [9], sparse coding (SC) and discriminating sparse coding (discSC) [8]. As outlined by [32] – there are two protocols for evaluation. In the first one (called 'training), a portion of the data from every household is used as training samples and rest (from those households) is used for prediction; this is the easier of the two protocols. In the second mode, the data from four households are used for training and the remaining one is used for prediction (called 'testing'); this is a more challenging problem. In this work, we carry out experiments on the more challenging problem, i.e. testing protocol.

The results are shown in Table II. The SC and discSC yields the best results for 144 atoms. For our method (both greedy and exact) the number of atoms are 144-100-80 in three layers. The table shows that our method is considerably superior compared to all other disaggregation techniques. The results are as expected. Results from discriminative sparse coding is slightly better than shallow sparse coding, but it worse compared to ours. The improvement from our greedy technique is decent, but it is not the best. The results obtained from our proposed exact solution yields the best results.

TABLE II
COMPARING DISAGGREGATION ACCURACIES FOR REDD

| House (tested on) | FHMM | SC | discSC | PED | Proposed (Greedy) | Proposed (Exact) |
|---|---|---|---|---|---|---|
| 1 | 46.6 | 57.17 | 58.11 | 46.0 | 60.76 | 64.26 |
| 2 | 50.8 | 65.42 | 68.25 | 49.2 | 71.05 | 74.93 |
| 3 | 33.3 | 41.06 | 42.40 | 31.7 | 43.50 | 48.26 |
| 4 | 52.0 | 60.25 | 73.76 | 50.9 | 76.75 | 79.02 |
| 6 | 55.7 | 58.06 | 53.93 | 54.5 | 61.71 | 64.19 |
| **Aggregate** | **47.7** | **56.39** | **59.29** | **46.5** | **62.75** | **66.13** |

TABLE III
COMPARING DISAGGREGATION ACCURACIES FOR PECAN STREET

| House (tested on) | FHMM | SC | discSC | PED | Proposed (Greedy) | Proposed (Eaxct) |
|---|---|---|---|---|---|---|
| 1 | 75.55 | 89.43 | 90.53 | 75.96 | 92.96 | 94.09 |
| 2 | 42.99 | 65.34 | 66.90 | 43.57 | 74.94 | 79.20 |
| 3 | 64.13 | 81.50 | 82.02 | 66.21 | 83.64 | 87.82 |
| 4 | 51.56 | 61.79 | 71.19 | 52.75 | 74.70 | 79.62 |
| 5 | 52.20 | 53.49 | 62.14 | 52.69 | 62.50 | 70.05 |
| 6 | 10.00 | 54.62 | 54.68 | 13.92 | 52.92 | 60.36 |
| 7 | 53.75 | 49.03 | 54.61 | 55.06 | 60.44 | 67.84 |
| 8 | 32.94 | 51.91 | 52.85 | 33.94 | 60.66 | 66.92 |
| 9 | 75.50 | 74.27 | 75.35 | 75.06 | 77.40 | 80.40 |
| 10 | 46.26 | 56.28 | 63.34 | 48.38 | 67.25 | 71.06 |
| 11 | 33.05 | 53.59 | 59.30 | 33.69 | 67.37 | 72.30 |
| 12 | 44.12 | 65.79 | 69.20 | 45.97 | 71.75 | 75.21 |
| 13 | 50.25 | 62.97 | 69.63 | 51.11 | 74.80 | 77.34 |
| 14 | 70.79 | 82.79 | 84.67 | 72.52 | 87.30 | 90.86 |
| 15 | 50.93 | 60.73 | 61.21 | 50.62 | 61.98 | 69.51 |
| 16 | 74.45 | 85.51 | 86.84 | 75.82 | 88.78 | 90.11 |
| 17 | 90.15 | 84.94 | 85.64 | 89.91 | 81.12 | 83.40 |
| 18 | 57.93 | 75.28 | 75.86 | 58.90 | 77.68 | 81.26 |
| 19 | 45.74 | 55.67 | 58.93 | 47.00 | 61.90 | 67.89 |
| 20 | 48.06 | 59.40 | 64.73 | 48.81 | 69.23 | 74.37 |
| 21 | 57.87 | 56.58 | 58.67 | 57.03 | 60.73 | 66.80 |
| 22 | 35.67 | 50.70 | 52.11 | 38.60 | 48.14 | 56.76 |
| 23 | 68.75 | 81.30 | 84.28 | 71.26 | 87.69 | 90.09 |
| 24 | 62.43 | 75.14 | 78.73 | 65.99 | 85.85 | 89.28 |
| 25 | 39.44 | 49.76 | 50.20 | 37.59 | 51.89 | 58.23 |
| 26 | 31.94 | 49.97 | 51.49 | 32.60 | 53.06 | 59.31 |
| 27 | 42.68 | 45.40 | 50.54 | 43.11 | 55.50 | 60.75 |
| 28 | 68.07 | 77.39 | 78.31 | 69.07 | 79.63 | 84.08 |
| 29 | 31.00 | 55.65 | 55.65 | 31.00 | 57.11 | 66.02 |
| 30 | 35.75 | 53.09 | 55.68 | 38.85 | 55.18 | 63.96 |
| 31 | 38.81 | 52.09 | 52.92 | 40.03 | 51.44 | 59.82 |
| 32 | 47.24 | 63.95 | 67.30 | 59.92 | 71.79 | 75.60 |
| 33 | 71.00 | 66.88 | 68.69 | 67.06 | 67.25 | 69.22 |
| 34 | 31.37 | 48.47 | 50.37 | 33.92 | 49.74 | 58.31 |
| 35 | 45.36 | 48.95 | 51.10 | 45.90 | 58.74 | 63.50 |
| 36 | 26.89 | 44.87 | 49.95 | 30.13 | 52.02 | 58.34 |
| 37 | 30.73 | 50.68 | 54.51 | 38.71 | 59.42 | 64.31 |
| 38 | 38.28 | 60.04 | 61.92 | 41.09 | 62.85 | 65.55 |
| 39 | 63.95 | 73.79 | 76.91 | 64.06 | 83.15 | 85.82 |
| 40 | 47.32 | 52.86 | 53.25 | 50.09 | 51.11 | 61.79 |
| 41 | 47.51 | 46.19 | 50.76 | 55.03 | 53.10 | 62.06 |
| 42 | 51.10 | 61.91 | 65.63 | 51.85 | 68.97 | 72.56 |
| 43 | 60.70 | 72.52 | 77.94 | 61.37 | 84.83 | 87.24 |
| 44 | 28.41 | 55.35 | 56.89 | 29.18 | 58.90 | 65.32 |
| 45 | 56.53 | 78.47 | 81.79 | 58.51 | 84.66 | 87.09 |
| 46 | 35.16 | 49.17 | 54.55 | 39.06 | 61.89 | 69.74 |
| 47 | 41.75 | 71.46 | 73.67 | 49.38 | 72.67 | 76.77 |
| **Aggregate** | **49.07** | **62.06** | **64.96** | **50.90** | **67.58** | **72.72** |

TABLE IV
NORMALIZED ERROR FOR COMMON DEVICES

| Appliance | FHMM | SC | discSC | PED | Proposed (Greedy) | Proposed (Exact) |
|---|---|---|---|---|---|---|
| AC | 3.16 | 0.90 | 0.70 | 2.52 | 0.89 | 0.80 |
| Dryer | 51.47 | 16.57 | 2.04 | 35.69 | 1.11 | 1.02 |
| Dishwasher | 6.48 | 4.23 | 1.25 | 6.08 | 0.66 | 0.62 |
| Microwave | 4.96 | 4.55 | 0.84 | 4..3 | 0.76 | 0.70 |
| Furnace | 0.89 | 0.79 | 0.63 | 0.93 | 0.58 | 0.55 |
| Fridge | 2722.8 | 916.53 | 516.3 | 986.30 | 490.56 | 401.78 |
| Washer | 21.80 | 8.75 | 0.93 | 19.62 | 0.59 | 0.55 |

*B. REDD Dataset*

We conduct this experiment on a subset of Dataport dataset available in NILMTK (non-intrusive load monitoring toolkit) format, which contains 1 minute circuit level and building level electricity data from 240 houses. The data set contains per minute readings from 18 different devices: air conditioner, kitchen appliances, electric vehicle, and electric hot tub heater, electric water heating appliance, dish washer, spin dryer, freezer, furnace, microwave, oven, electric pool heater, refrigerator, sockets, electric stove, waste disposal unit, security alarm and washer dryer. We are assigning about 80% of the homes to the training set and the remaining 20% of the homes to the test set. To prepare training and testing data, aggregated and sub-metered data are averaged over a time period of 10 minutes. This is the usual protocol to carry out experiments on the Pecan street dataset. Each training sample contains power consumed by a particular device in one day while each testing sample contains total power consumed in one day in particular house.

The number of atoms for different techniques remain the same as before. The results are shown in Table III. The conclusion remains the same as before. Our method outperforms other techniques by a wide margin. The interesting observation here is that by deep sparse coding, we are able to get significantly larger improvement on homes where the disaggregation accuracy was previously lower, e.g. 6-8, 15, 29 etc.

For the Pecan Street dataset, we also study the variation of performance with respect to different electrical appliances. The metric used here is Normalized Error. The results are shown in Table IV. The results show that our proposed method yields the best disaggregation in terms of normalised error for every device. FHMM and PED yields significantly worse results. Sparse coding and discriminating sparse coding yield reasonably good results but is worse than our proposed deep sparse coding.

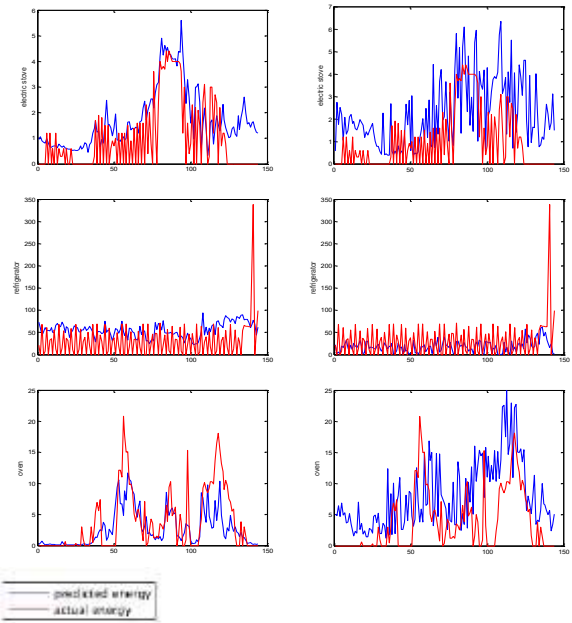

Fig. 10. Energy Disaggregation: Qualitative Look. Left – Proposed Greedy Method; Right – discSC [8].

To visually show the disaggregation results for the Pecan Street dataset, some samples are shown in the Fig. 10. The red plot shows the actual energy consumed and the blue plot the predicted energy. One can see that even with our proposed greedy method, the estimated and the actual values are close, while results from [8] are considerably off.

## V. Conclusion

The problem of energy disaggregation has attracted attention of the machine learning community in recent times. Broadly speaking it is a single channel blind source separation problem. The current trend in disaggregation is to learn a dictionary for each device by taking into account some prior information regarding the same. The learnt dictionaries are used for blind source separation leading to disaggregation.

So far all the techniques for learning the basis are shallow, i.e. a single layer of dictionary is learnt for each device. Given the success of deep learning in various machine learning applications, we propose to learn multiple layers of dictionaries for each device. We call this – deep sparse coding. Experimental results on two benchmark datasets show that our proposed method is always better than the state-of-the-art methods in energy disaggregation.

The shortcoming of our work (and all other studies based on sparse coding / dictionary learning) is that, it cannot be used for real-time disaggregation. If such be the need, HMM based techniques [36] would be more suitable.

Prior studies [8, 9] have shown that better results can be obtained (for shallow techniques) when further assumptions regarding the device are made. In future we would like to incorporate it into our deep learning framework and hope to improve the results even further. On a practical front, we would like to see if our technique can be used to disaggregate specific loads, e.g. one may be interested in consumption of heavy loads such as AC [35, 36].

## Acknowledgement

We would like to thank Department of Electronics and Information Technology (Government of India) for the Grant Number ITRA/15(57)/Mobile/HumanSense/01.